\documentclass[12pt,onecolumn,technote]{IEEEtran}

\usepackage{mathrsfs}
\usepackage{txfonts}
\usepackage [dvips] {graphicx}
\usepackage{graphicx}
\usepackage{caption}
\usepackage{subcaption}
\usepackage[ruled]{algorithm2e}
\begin{document}

\title{Compressed Sensing of Multi-Channel EEG Signals: The Simultaneous Cosparsity and Low Rank Optimization}

\author{Yipeng~Liu*,~\IEEEmembership{Member,~IEEE},
        Maarten~De~Vos,~\IEEEmembership{Member,~IEEE},
        and~Sabine~Van~Huffel,~\IEEEmembership{Fellow,~IEEE}

\thanks{This work was supported by FWO of Flemish Government: G.0108.11 (Compressed Sensing); Belgian Federal Science Policy Office:  IUAP P7/19/ (DYSCO, `Dynamical systems, control and optimization', 2012-2017); ERC Advanced Grant: BIOTENSORS (339804). \emph{Asterisk indicates corresponding author}.
}

\thanks{*Yipeng Liu is with School of Electronic Engineering / Center for Robotics / Center for Information in BioMedicine, University of Electronic Science and Technology of China (UESTC), Xiyuan Avenue 2006, Western High-Tech Zone, Chengdu, 611731, China. Most of this work has been done when he was with ESAT-STADIUS Division / iMinds Medical IT Department, Dept. of Electrical Engineering, University of Leuven, Kasteelpark Arenberg 10, box 2446, 3001 Leuven, Belgium. (e-mail: yipengliu@uestc.edu.cn) }%
\thanks{Maarten De Vos is with Institute of Biomedical Engineering, Department of Engineering, University of Oxford, Oxford, United Kingdom. }
\thanks{Sabine Van Huffel is with ESAT-STADIUS Division / iMinds Medical IT Department, Dept. of Electrical Engineering, University of Leuven, Kasteelpark Arenberg 10, box 2446, 3001 Leuven, Belgium. }%
\thanks{Copyright (c) 2014 IEEE. Personal use of this material is permitted. However, permission to use this material for any other purposes must be obtained from the IEEE by sending an email to pubs-permissions@ieee.org. }%
}

\markboth{Accepted by IEEE Transactions on Biomedical Engineering}%
{Shell \MakeLowercase{\textit{et al.}}: Bare Demo of IEEEtran.cls for Journals}

\maketitle

\begin{abstract}
\emph{Goal}: This paper deals with the problems that some EEG signals have no good sparse representation and single channel processing is not computationally efficient in compressed sensing of multi-channel EEG signals.
\emph{Methods}: An optimization model with L0 norm and Schatten-0 norm is proposed to enforce cosparsity and low rank structures in the reconstructed multi-channel EEG signals. Both convex relaxation and global consensus optimization with alternating direction method of multipliers are used to compute the optimization model.
\emph{Results}: The performance of multi-channel EEG signal reconstruction is improved in term of both accuracy and computational complexity.
\emph{Conclusion}: The proposed method is a better candidate than previous sparse signal recovery methods for compressed sensing of EEG signals.
\emph{Significance}: The proposed method enables successful compressed sensing of EEG signals even when the signals have no good sparse representation. Using compressed sensing would much reduce the power consumption of wireless EEG system.

\end{abstract}


\IEEEpeerreviewmaketitle

\section{Introduction}
\label{sec1}

\IEEEPARstart{W}{ireless} body sensor networks take spatially distributed sensors to acquire physiological signals, and transmit them over wireless links to a central unit for signal processing \cite{bachmann2012low}. The electroencephalogram (EEG) signal is one of the most frequently used biomedical signals. It has important applications in medical healthcare, brain computer interfacing (BCI), and so on \cite{devos2014mobile}. Continuous EEG monitoring usually requires large amount of data to be sampled and transmitted, which leads to large size of batteries. The recording unit of the wireless portable EEG systems is powered with batteries, and the physical size of the batteries sets the overall device size and operational lifetime. A physically too large device would not be portable; and excessive battery power consumption would make the long time wireless recording very hard \cite{debener2012how}  \cite{abdulghani2012compressive} \cite{zhang2013compressed}.

Compressed sensing (CS) was proposed to deal with this challenge. Rather than first sample the analog signal at Nyquist rate and discard most in the compression, it directly acquires the digital compressed measurements at a lower sampling rate, and recovers the digital signals by nonlinear algorithms from the compressed measurements \cite{eldar2012compressed}. CS relies on the assumption that the signal vector \textbf{x} is compressed by a random matrix $ {\mathbf{\Phi }} \in {\mathbb{R}^{M \times N}} $  (measurement or sampling matrix) in discrete form as \cite{eldar2012compressed} \cite{becker2011practical}:
\begin{equation}
\label{eq1 sampling model}
{\bf{y}} = {\bf{\Phi x}},
\end{equation}
where \textbf{y} is the random sub-Nyquist compressed measurement. Here $ M \ll N $, which means that it is sampled at a greatly reduced rate. If \textbf{x} is sparse, its recovery only requires the compressed signal \textbf{y} and the sampling matrix $ \Phi $. If it is not sparse, the signal \textbf{x} should be represented (transformed) using a representation matrix (dictionary) $ {\mathbf{\Psi }} \in {\mathbb{R}^{N \times P}} $ with $ N \leqslant P $ and a sparse vector $ {\mathbf{\theta }} \in {\mathbb{R}^{P \times 1}} $ with most of its entries zero or almost zero as:
\begin{equation}
\label{eq1 sparse signal model}
{\mathbf{x}} = {\mathbf{\Psi \theta }}.
\end{equation}
With the compressed measurement \textbf{y}, sampling matrix $ {\mathbf{\Phi }} $ and dictionary $ {\bf{\Psi }} $, we can recover \textbf{x} by (\ref{eq1 sparse signal model}) after computing $ \theta $ by:
\begin{equation}
\label{eq1 sythesis L0 optimization}
\begin{array}{c}
\mathop {{\mathop{\rm minimize}\nolimits} }\limits_{\bf{\theta }} {\left\| {\bf{\theta }} \right\|_0}\\
{\rm{subject~to~~ }}{\bf{y}} = {\bf{\Phi \Psi \theta }}
\end{array},
\end{equation}
where $ {\left\| {\bf{\theta }} \right\|_0} $ is the pseudo-$ \ell_0 $ norm which counts the number of nonzero entries, i.e. $ {\left\| {\bf{\theta}} \right\|_0} = \# \{ {\theta_n} \ne 0,~{\rm{ }}n = 1,2, \cdots ,N\} $. The signal \textbf{x} is called \emph{K}-sparse when the number of nonzero entries is \emph{K}. Most of the current methods for biomedical signal recovery from compressed samples are based on the solution of the $ \ell_0 $ programming problem (\ref{eq1 sythesis L0 optimization}), such as, basis pursuit (BP), orthogonal matching pursuit (OMP), iterative hard thresholding (IHT), etc \cite{abdulghani2012compressive} \cite{tropp2010computational} \cite{mamaghanian2011compressed}. Besides, \cite{zhang2013compressed} found that some EEG signals are not sparse in any sparse transformed domains, and proposed to exploit block-sparsity by block sparse Bayesian learning (BSBL) to recover EEG signals \cite{zhang2013compressed}.

Contrary to the traditional sparse or block-sparse signal model, the cosparse signal model uses an analysis operator multiplying the measurement to produce a sparse vector \cite{nam2013cosparse}:
\begin{equation}
\label{eq2.1 cosparse signal model}
{\mathbf{\rho }} = {\mathbf{\Omega x}},
\end{equation}
where $ {\mathbf{\Omega }} \in {\mathbb{R}^{Q \times N}} $ is the cosparse representation matrix (analysis dictionary) with $ N \leqslant Q $, and $ {\mathbf{\rho }} \in {\mathbb{R}^{Q \times 1}} $ is the cosparse vector if most of its entries are nearly zero. Several sufficient conditions theoretically guarantee the successful recovery of the cosparse signal  from the compressed measurement, such as the restricted isometry property adapted to the dictionary (D-RIP), restricted orthogonal projection property (ROPP), etc \cite{nam2013cosparse} \cite{candes2011compressed} \cite{peleg2013performance}. When $ N = P $, an equivalent cosparse signal model to the sparse signal model can be found by letting ${\bf{\Omega }}{\rm{ = }}{{\bf{\Psi }}^{{\rm{ - 1}}}} $; but there is no such an equivalent when $  N < P $.  The traditional sparse synthesis model puts an emphasis on the non-zeros of the sparse vector $ \mathbf{\theta} $, but the cosparse analysis model draws its strength from the zeros of the analysis vector $ \mathbf{\rho } $.

%

The cosparse signal recovery has some unique advantages in CS based EEG systems. First, the sparse signal recovery (\ref{eq1 sythesis L0 optimization}) gets the best estimate of the sparse vector $ \mathbf{\theta} $; but the cosparse signal recovery (\ref{eq3 analysis L0 optimization}) gets the EEG signal's best estimate directly. Second, theoretically the sparse signal recovery (\ref{eq1 sythesis L0 optimization}) requires the columns of the representation matrix $ \mathbf{\Psi } $ to be incoherent, but the cosparse way (\ref{eq3 analysis L0 optimization}) allows the coherence of the cosparse representation matrix $ \mathbf{\Omega } $, which can result in super resolution of the EEG signal estimate \cite{candes2011compressed}. Third, the EEG signal can hardly be sparsely represented \cite{zhang2013compressed}. However, data analysis shows that the EEG signals are approximately piecewise linear \cite{yan2013an}, as shown in Fig. \ref{fig101}, which implies the signal fits the cosparse signal model (\ref{eq2.1 cosparse signal model}) well with the 2nd order difference matrix as the cosparse analysis dictionary. Therefore, the cosparse signal recovery should be more appropriate for CS of EEG signals.

Since nearly all types of EEG systems have multiple channels, it can be taken for granted that it is better to jointly process the multi-channel EEG signals. \cite{durka2005multichannel} proposed to jointly process multi-channel EEG signals by allowing slightly different phases of the dictionaries in different channels. Another classical way assumes that multiple channels share a similar support of sparse vector. This generalizes the single measurement vector (SMV) problem straightforwardly to a multiple measurement vector (MMV) problem \cite{cotter2005sparse} \cite{zhang2014spatiotemporal}. \cite{fauvel2014energy} proposed to incorporate preprocessing and entropy coding in the sampling to reduce the redundance in correlated multi-channel signals, but the added preprocessing and encoder would increase the power consumption in EEG sampling  \cite{abdulghani2012compressive}; and the procedure can hardly be realized for analog signals, which implies the analog EEG signals should be sampled at Nyquist sampling rate in the beginning. To compress the multi-channel EEG signals from the complete digital measurement, \cite{srinivasan2013multichannel} used a wavelet-based volumetric coding method, while \cite{dauwels2013near} exploited the low rank structure in matrix/tensor form and achieved better performance.

Since most of the multi-channel EEG signals are more or less correlated with each other, the low rank structure based compression method motivates the use of low rank data structure in CS of multi-channel EEG signals too. The multi-channel EEG signals are put columnwise into a matrix. Our EEG data analysis finds that the newly formed EEG data matrix has only a few nonzero singular values.

In this paper, the 2nd-order difference matrix is chosen to be the cosparse analysis dictionary, which tries to enforce the approximate piecewise linear structure. Exploiting additionally the low rank structure, we can further enhance the signal recovery performance by exploiting the cosparsity of single channel EEG signals and the low rank property of multi-channel EEG signals simultaneously in the framework of multi-structure CS. The $ \ell_0 $ norm and Schatten-0 norm based optimization model is used to encourage cosparsity and low rank structure in the reconstructed signals. Two methods are proposed to solve the multi-criterion optimization problem. One relaxes it to a convex optimization; and the other one transforms it into a global consensus optimization problem. The alternating direction method of multipliers (ADMM) is used to solve it efficiently. The convergence and computational complexity are briefly analyzed. In numerical experiments, a group of real-life EEG data is used to test the algorithms' performance of both single-channel and multi-channel EEG signal recovery methods. Numerical results show that the cosparse signal recovery method and simultaneous cosparsity and low-rank (SCLR) optimization achieve the best performance in term of mean squared error (MSE) and mean cross-correlation (MCC) in single channel and multi-channel EEG signal recovery respectively.

%
%


The rest of the paper is organized as follows. Section \ref{sec3} presents an optimization model to exploit both cosparsity and low rank data structures to recover the EEG signals. In Section \ref{sec4}, two methods are given to solve the optimization problem, i.e. convex relaxation and  alternating direction method of multipliers (ADMM). In Section \ref{sec5}, numerical experiments are used to demonstrate the proposed methods' performance improvement. Section \ref{sec6} draws the conclusion.

\section{Simultaneous Cosparsity and Low Rank Optimization Model}
\label{sec3}

The optimization model for cosparse signal recovery can be formulated as \cite{nam2013cosparse}:
\begin{equation}
\label{eq3 analysis L0 optimization}
\begin{array}{c}
\mathop {{\mathop{\rm minimize}\nolimits} }\limits_{\bf{x}} {\left\| {{\bf{\Omega x}}} \right\|_0}\\
{\rm{subject~to~~ }}{\bf{y}} = {\bf{\Phi x}}
\end{array}.
\end{equation}
Here we call (\ref{eq3 analysis L0 optimization}) the analysis L0 optimization. When the EEG system records \emph{R} channels simultaneously, the extension of analysis L0 optimization to multi-channel data is:
\begin{equation}
\label{eq3 analysis L0 optimization: matrix}
\begin{array}{c}
\mathop {{\mathop{\rm minimize}\nolimits} }\limits_{\bf{X}} {\left\| {{\mathop{\rm vec}\nolimits} \left( {{\bf{\Omega X}}} \right)} \right\|_0}\\
{\rm{subject~to~~ }}{\bf{Y}} = {\bf{\Phi X}}
\end{array},
\end{equation}
where $ {\mathbf{X}} \in {\mathbb{R}^{N \times R}} $, and $ {\mathop{\rm vec}\nolimits} ({\bf{X}}) $ puts all the columns of \textbf{X} into one column vector sequentially. A series of solvers are summarized in \cite{nam2013cosparse}.

Reconstructing the EEG matrix from the compressed measurements by exploiting the low rank structure can be formulated as:
\begin{equation}
\label{eq3 Schatten0 optimization}
\begin{array}{c}
\mathop {{\mathop{\rm minimize}\nolimits} }\limits_{\bf{X}} {\left\| {\bf{X}} \right\|_{{\rm{Schatten}} - 0}}\\
{\rm{subject~to~~ }}{\bf{Y}}{\rm{ = }}{\bf{\Phi X}}
\end{array},
\end{equation}
where $ {\left\| {\bf{X}} \right\|_{{\rm{Schatten}} - 0}} $  is the Schatten-0 norm which counts the number of the nonzero singular values of \textbf{X} \cite{rohde2011estimation}. A variety of methods to solve it can be found in \cite{recht2010guaranteed}.

Motivated by the fact that many EEG signals have both cosparsity and low rank structure, we propose to simultaneously exploit these two data structures in multi-channel EEG signal reconstruction from the compressed measurement. Both $ \ell_0 $ norm and Schatten-0 norm based constraints are used in the optimization model. Combining with the linear data fitting constraint, we can formulate the simultaneous cosparsity and low rank (SCLR) optimization model as follows:
\begin{equation}
\label{eq3 L0-Schatten0 optimization}
\begin{array}{c}
\mathop {{\rm{minimize}}}\limits_{\bf{X}} {\left\| {{\mathop{\rm vec}\nolimits} \left( {\bf{\Omega}} {\bf{X}} \right)} \right\|_0} + {\left\| {\bf{X}} \right\|_{{\rm{Schatten}} - 0}}\\
{\rm{subject~to~~ }}{\bf{Y}}{\rm{ = }}{\bf{\Phi X}}
\end{array}.
\end{equation}

\section{solutions}
\label{sec4}

\subsection{Convex relaxation}
\label{sec4.1}
To solve the SCLR optimization (\ref{eq3 L0-Schatten0 optimization}), one classical way relaxes the nonconvex $ \ell_0 $ norm and Schatten-0 norm into convex $ \ell_1 $ norm and Schatten-1 norm respectively, where the $ \ell_1 $ norm sums all the absolute values of the entries, i.e. $ {\left\| {\bf{x}} \right\|_1} = \sum\nolimits_{n = 1}^N {\left| {{x_n}} \right|} $. The Schatten-1 norm is called nuclear norm too, and sums all the singular values of the data matrix, i.e. $ {\left\| {\bf{X}} \right\|_{{\rm{Schatten}} - 1}} = {\left\| {\bf{X}} \right\|_*} = \sum\nolimits_{n = 1}^{\min (N,P)} {{\sigma _n}}  $. The newly formed convex simultaneous cosparsity and low rank (CSCLR) optimization model can be formulated as:
\begin{equation}
\label{eq4.1 L1-Schatten1 optimization}
\begin{array}{c}
\mathop {{\rm{minimize}}}\limits_{\bf{ X}} {\left\| {{\mathop{\rm vec}\nolimits} \left( {\bf{ \Omega  X}} \right)} \right\|_1} + {\left\| {\bf{X}} \right\|_*}\\
{\rm{subject~to~~ }}{\bf{Y}}{\rm{ = }}{\bf{\Phi X}}
\end{array}.
\end{equation}

Similarly to the reformulation from $ {\rm{minimiz}}{{\rm{e}}_{\bf{x}}}{\left\| {\bf{x}} \right\|_1} $ to  $ {\rm{minimiz}}{{\rm{e}}_{{\bf{x}},{\bf{e}} \succ 0}}~{{\bf{1}}^T}{\bf{e}},~{\rm{subject~ to~ }} - {\bf{e}} \prec {\bf{x}} \prec {\bf{e}} $ due to the definition of the $ \ell_1 $ norm, we can re-formulate the $ \ell_1 $ norm minimization into its equivalent linear programming in (\ref{eq4.1 L1-Schatten1 optimization}) \cite{boyd2009convex}. By introduction of new nonnegative variables \textbf{e} and \emph{f}, (\ref{eq4.1 L1-Schatten1 optimization}) can be expressed as:
\begin{equation}
\label{eq4.1 L1-Schatten1 optimization: 2}
\begin{array}{r}
\mathop {{\rm{minimize}}}\limits_{{\bf{X}},{\bf{e}} \succ 0,~f \ge 0} {\rm{ }}{{\bf{1}}^T}{\bf{e}} + f\\
{\rm{subject~ to~~ }}{\bf{Y}}{\rm{ = }}{\bf{\Phi X}}\\
{\left\| {\bf{X}} \right\|_*} \le f\\
 - {\bf{e}} \prec {\mathop{\rm vec}\nolimits} \left( {{\bf{\Omega X}}} \right) \prec {\bf{e}}
\end{array},
\end{equation}
where $ {\mathbf{1}} \in {\mathbb{R}^{QR \times 1}} $ is a column vector with all the entries being 1.

The nuclear norm constraint can be replaced by its linear matrix inequality (LMI) equivalent; and the approximation constraints can also be expressed via LMIs using Schur complements \cite{vandenberghe1996semidefinite}. The obtained optimization model is:
\begin{equation}
\label{eq4.1 L1-Schatten1 optimization: 3}
\begin{array}{r}
\mathop {{\rm{minimize}}}\limits_{{\bf{X}},{\bf{e}} \succ 0,f \ge 0} {\rm{ }}{{\bf{1}}^T}{\bf{e}} + 2f\\
{\rm{subject~ to~~ }}{\bf{Y}}{\rm{ = }}{\bf{\Phi X}}\\
 - {\bf{e}} \prec {\mathop{\rm vec}\nolimits} \left( {{\bf{\Omega X}}} \right) \prec {\bf{e}}\\
\left[ {\begin{array}{*{20}{c}}
{\bf{A}}&{\bf{X}}\\
{{{\bf{X}}^T}}&{\bf{B}}
\end{array}} \right] \ge 0\\
{\rm{Tr}}\left( {\bf{A}} \right){\rm{ + Tr}}\left( {\bf{B}} \right) < f
\end{array},
\end{equation}
where $ {\bf{{ A}}} = {{\bf{{ A}}}^T} $  and $ {\bf{B}} = {{\bf{B}}^T} $ are new variables. (\ref{eq4.1 L1-Schatten1 optimization: 3}) is a semi-definite programming (SDP) which can be solved by interior-point method \cite{boyd2009convex} \cite{vandenberghe1996semidefinite}. The software CVX can compute the solution in this way \cite{grant2012cvx}.

\subsection{ADMM}

Besides the classical SDP, another method, called ADMM, can be used to solve the SCLR optimization \cite{boyd2011distributed}. With individual constraints on the same variables in each constraint,  (\ref{eq4.1 L1-Schatten1 optimization}) can be rewritten into a global consensus optimization with local variables $ {{\bf{X}}_i} $, \emph{i} = 1,2 and a common global variable \textbf{X} as:
\begin{equation}
\label{eq4.2 consensus optimization}
\begin{array}{c}
\mathop {{\rm{minimize}}}\limits_{{{\bf{X}}_1},{{\bf{X}}_2},{\bf{X}}} {\left\| {{\mathop{\rm vec}\nolimits} \left( {{\bf{\Omega }}{{\bf{X}}_1}} \right)} \right\|_1} + {\left\| {{{\bf{X}}_2}} \right\|_*}\\
{\rm{subject~  to~~  }}{\bf{X}}{\rm{ = }}{{\bf{X}}_1};{\rm{ }}{\bf{X}}{\rm{ = }}{{\bf{X}}_2};{\bf{Y}} = {\bf{\Phi X}}
\end{array}.
\end{equation}
Here the new constraints are that all the local variables should be equal. It is equivalent to:
\begin{equation}
\label{eq4.2 consensus optimization: compact form}
\begin{array}{c}
\mathop {{\rm{minimize}}}\limits_{{{\bf{X}}_1},{{\bf{X}}_2},{\bf{X}}} {\left\| {{\mathop{\rm vec}\nolimits} \left( {{\bf{\Omega }}{{\bf{X}}_1}} \right)} \right\|_1} + {\left\| {{{\bf{X}}_2}} \right\|_*}\\
{\rm{subject~ to~~  }}{{\bf{\bar Y}}}{\rm{ = }}{\bf{\bar \Phi }}{{\bf{X}}_1};{\rm{ }}{{\bf{\bar Y}}}{\rm{ = }}{\bf{\bar \Phi }}{{\bf{X}}_2}
\end{array},
\end{equation}
where
\begin{equation}
\label{eq4.2 consensus optimization: compact form 1}
{\bf{ \bar Y}} = \left[ {\begin{array}{*{20}{c}}
{{\bf{ Y}}}\\
{\bf{X}}
\end{array}} \right],
\end{equation}

\begin{equation}
\label{eq4.2 consensus optimization: compact form 2}
{\bf{\bar \Phi }} = \left[ {\begin{array}{*{20}{c}}
{\bf{\Phi }}\\
{\bf{I}}
\end{array}} \right].
\end{equation}

The corresponding augmented Lagrangian of (\ref{eq4.2 consensus optimization: compact form}) is:
\begin{equation}
\label{eq4.2 augmented Lagrangian}
\begin{array}{l}
{L_\rho }\left( {{ {  \bf{ X}}_1},{{\bf{X}}_2};{{\bf{Z}}_1},{{\bf{Z}}_2}} \right) = {\left\| {{\mathop{\rm vec}\nolimits} \left( { \bf{ \Omega} {{\bf{X}}_1}} \right)} \right\|_1} + {\left\| {{{\bf{X}}_2}} \right\|_*}\\
 + {\mathop{\rm vec}\nolimits} {\left( {{{\bf{Z}}_1}} \right)^T}{\mathop{\rm vec}\nolimits} \left( {{{\bf{\bar Y}}} - {\bf{\bar \Phi }}{{\bf{X}}_1}} \right) + {\mathop{\rm vec}\nolimits} {\left( {{{\bf{Z}}_2}} \right)^T}{\mathop{\rm vec}\nolimits} \left( {{{\bf{\bar Y}}} - {\bf{\bar \Phi }}{{\bf{X}}_2}} \right)\\
 + \frac{\rho }{2}\left\| {{{\bf{\bar Y}}} - {\bf{\bar \Phi }}{{\bf{X}}_1}} \right\|_F^2 + \frac{\rho }{2}\left\| {{{\bf{\bar Y}}} - {\bf{\bar \Phi }}{{\bf{X}}_2}} \right\|_F^2
\end{array},
\end{equation}
where $ \rho >0 $, $ {{\mathbf{Z}}_1} $ and $ {{\mathbf{Z}}_2} $ are dual variables. The resulting ADMM algorithm in the scaled dual form is the following
\begin{equation}
\label{eq4.2 admm:1}
{\bf{X}}_1^{t + 1}: = \mathop {\arg \min }\limits_{{{\bf{X}}_1}} \left( {{{\left\| {{\mathop{\rm vec}\nolimits} \left( {{\bf{\Omega }}{{\bf{X}}_1}} \right)} \right\|}_1} + \frac{\rho }{2}\left\| {{{\bf{\bar Y}}} - {\bf{\bar \Phi }}{{\bf{X}}_1} + {\bf{\bar \Phi }}{\bf{U}}_1^t} \right\|_F^2} \right),
\end{equation}

\begin{equation}
\label{eq4.2 admm:2}
{\bf{X}}_2^{t + 1}: = \mathop {\arg \min }\limits_{{{\bf{X}}_2}} \left( {{{\left\| {{{\bf{X}}_2}} \right\|}_*} + \frac{\rho }{2}\left\| {{{\bf{ \bar Y}}} - {\bf{\bar \Phi }}{{\bf{X}}_2} + {\bf{\bar \Phi }}{\bf{U}}_2^t} \right\|_F^2} \right),
\end{equation}

\begin{equation}
\label{eq4.2 admm:3}
{{\bf{X}}^{t + 1}} = \frac{1}{2}\left( {{\bf{X}}_1^{t + 1} + {\bf{X}}_2^{t + 1}} \right),
\end{equation}

\begin{equation}
\label{eq4.2 admm:4}
\begin{array}{l}
{\bf{U}}_1^{t + 1} = {\bf{U}}_1^t + \left( {{\bf{X}}_1^{t + 1} - {{\bf{X}}_1^t}} \right)\\
{\bf{U}}_2^{t + 1} = {\bf{U}}_2^t + \left( {{\bf{X}}_2^{t + 1} - {{\bf{X}}_2^t}} \right)
\end{array},
\end{equation}
where $ {{\mathbf{U}}_1} = {1 \mathord{\left/
 {\vphantom {1 \rho }} \right.
 \kern-\nulldelimiterspace} \rho }{{\mathbf{Z}}_1} $ and $ {{\mathbf{U}}_2} = {1 \mathord{\left/
 {\vphantom {1 \rho }} \right.
 \kern-\nulldelimiterspace} \rho }{{\mathbf{Z}}_2} $  are scaled dual variables. In the proposed ADMM algorithm for SCLR optimization, two steps separately optimize over variables generally, i.e. updating the prime variables $ {{\mathbf{X}}_1} $ and $ {{\mathbf{X}}_2} $,  updating the scaled dual variables $ {{\mathbf{U}}_1} $ and $ {{\mathbf{U}}_2} $. In this iterative algorithm, the variables are updated in an alternating fashion.

For both (\ref{eq4.2 admm:1})  and (\ref{eq4.2 admm:2}), there are many computationally efficient algorithms \cite{nam2013cosparse} \cite{recht2010guaranteed}. For example, analysis L1 optimization,  greedy analysis pursuit (GAP) can be used to solve (\ref{eq4.2 admm:1}); to solve (\ref{eq4.2 admm:2}), SDP method or singular value thresholding (SVT) can be used. The solutions of (\ref{eq4.2 admm:3}) and (\ref{eq4.2 admm:4}) are straightforwardly easy. The ADMM for SCLR optimization is summarized in Algorithm 1.

\begin{algorithm}[htbp]
\label{alg1}
\small
\caption{ADMM for the SCLR optimization}
$\bullet$ Set $t := 0$, a small scalar ${\eta} > 0$, ${\bf{U}}_1^0$, ${\bf{U}}_2^0$, $ {{{\bf{X}}^0}} $, $ {T_{\max }} $, $ \rho $; \\

\Repeat{$\frac{{{{\left\| {{{\bf{X}}^{t + 1}} - {{\bf{X}}^t}} \right\|}_F}}}{{{{\left\| {{{\bf{X}}^{t + 1}}} \right\|}_F}{{\left\| {{{\bf{X}}^t}} \right\|}_F}}} \le \eta $ or $ t = {T_{\max }} $ }
{
$\bullet$ step 1:  update of the analysis L1 optimization: ${\bf{X}}_1^{t + 1}: = \mathop {\arg \min }\limits_{{{\bf{X}}_1}} \left( {{{\left\| {{\mathop{\rm vec}\nolimits} \left( {{\bf{\Omega }}{{\bf{X}}_1}} \right)} \right\|}_1} + \frac{\rho }{2}\left\| {{{{\mathbf{\bar Y}}}} - {\bf{\bar \Phi }}{{\bf{X}}_1} + {\bf{\bar \Phi }}{\bf{U}}_1^t} \right\|_F^2} \right)$ ;

$\bullet$step 2: update of the low rank optimization: ${\bf{X}}_2^{t + 1}: = \mathop {\arg \min }\limits_{{{\bf{X}}_2}} \left( {{{\left\| {{{\bf{X}}_2}} \right\|}_*} + \frac{\rho }{2}\left\| {{{{\mathbf{\bar Y}}}} - {\bf{\bar \Phi }}{{\bf{X}}_2} + {\bf{\bar \Phi }}{\bf{U}}_2^t} \right\|_F^2} \right)$;

$\bullet$step 3: update of the global variable: ${{\bf{X}}^{t + 1}} = \frac{1}{2}\left( {{\bf{X}}_1^{t + 1} + {\bf{X}}_2^{t + 1}} \right)$;

$\bullet$step 4: update of the dual variables: $\begin{array}{l}
{\bf{U}}_1^{t + 1} = {\bf{U}}_1^t + \left( {{\bf{X}}_1^{t + 1} - {{\bf{X}}_1^t}} \right)\\
{\bf{U}}_2^{t + 1} = {\bf{U}}_2^t + \left( {{\bf{X}}_2^{t + 1} - {{\bf{X}}_2^t}} \right)
\end{array}$;

$\bullet$ step 5: set $t := t + 1$;
}
$\bullet$ Algorithm ends and return ${{\bf{X}}^{t + 1}}$.
\end{algorithm}

A lot of convergence results exist for ADMM in the literature \cite{boyd2011distributed}. Generally, the convergence to optimum can be guaranteed when the epigraph of $ {g_i} $:

\begin{equation}
\label{eq4.2 epi}
{\mathop{\rm epi}\nolimits} {g_i} = \left\{ {\left( {{\bf{X}},\varepsilon } \right) \ |~{g_i}\left( {\bf{X}} \right) \le \varepsilon },~i = 1, 2. \right\}
\end{equation}
is a closed nonempty convex set, where $ {g_1}({\bf{X}}) = {\left\| {{\mathop{\rm vec}\nolimits} \left( {{\bf{\Omega X}}} \right)} \right\|_1} $, $ {g_2}({\bf{X}}) = {\left\| {\bf{X}} \right\|_*} $, and the unaugmented Lagrangian
\begin{equation}
\label{eq4.2 unaugmented Lagrangian}
\begin{array}{l}
{L_{\rho  = 0}}\left( {  {{\bf{X}}_1},{{\bf{X}}_2};{{\bf{Z}}_1},{{\bf{Z}}_2}} \right) = {\left\| {{\mathop{\rm vec}\nolimits} \left( \bf{\Omega} {{{\bf{X}}_1}} \right)} \right\|_1} + {\left\| {{{\bf{X}}_2}} \right\|_*}\\
 + {\mathop{\rm vec}\nolimits} {\left( {{{\bf{Z}}_1}} \right)^T}{\mathop{\rm vec}\nolimits} \left( {{{\bf{\bar Y}}} - {\bf{\bar \Phi }}{{\bf{X}}_1}} \right) + {\mathop{\rm vec}\nolimits} {\left( {{{\bf{Z}}_2}} \right)^T}{\mathop{\rm vec}\nolimits} \left( {{{\bf{\bar Y}}} - {\bf{\bar \Phi }}{{\bf{X}}_2}} \right)
\end{array}
\end{equation}
has a saddle point. The proof can be found in \cite{eckstein1992douglas}.

The ADMM decomposes the optimization model with multiple constraints into several ones with fewer constraints. There could be some fast algorithms for these new optimization models. Besides, it allows multiple steps in one iteration to be processed in parallel. With a multi-core processor, the computational time can be decreased. Previous experience shows that a few iterations will often produce acceptable results of practical use.

\section{Numerical Experiments}
\label{sec5}

To demonstrate the performance of the possible methods for EEG signal recovery from the compressed measurement, we perform two groups of numerical experiments. The details about the data materials and subjects are given in section \ref{sec5.0}. In section \ref{sec5.1}, we test the performance of two cosparse signal recovery methods for single-channel EEG signals in different kinds of situations, i.e. analysis L1 optimization and GAP. Some other algorithms are tested to make comparison, such as BSBL which is reported to be the best of all the current candidates for EEG signal recovery from compressed measurement \cite{zhang2013compressed}, and OMP which is a proper representative of the classical sparse signal recovery algorithms \cite{abdulghani2012compressive}. In section \ref{sec5.2}, a group of multi-channel EEG signals are recovered by the proposed algorithms for SCLR optimization, as well as simultaneous orthogonal matching pursuit (SOMP) \cite{tropp2006algorithms}, BSBL \cite{zhang2013compressed} \cite{zhang2014spatiotemporal}, and simultaneous greedy analysis pursuit (SGAP) \cite{avonds2014simultaneous}.

In all experiments, as argued by our analysis in section \ref{sec1}, the 2nd-order difference matrix is chosen to be the analysis dictionary for cosparse EEG signal recovery. The Gaussian matrix is chosen to be the sampling matrix for CS of EEG signals. The sparse dictionaries of OMP and SOMP are Daubechies wavelets \cite{abdulghani2012compressive}.

To measure the compression degree, the subsampling ratio (SSR) is defined as:
\begin{equation}
\label{eq5 SSR}
\mathop{\rm SSR} = \frac{M}{N} \times 100\%.
\end{equation}

To quantify the difference between high-dimensional values implied by the estimator and the true values of the quantity being estimated, two different evaluation functions are often used in EEG signal processing. One is the mean squared error (MSE) which measures the average of the squares of the errors. The error is the amount by which the value implied by the estimator differs from the quantity to be estimated. Here we can formulate it as:

\begin{equation}
\label{eq5 MSE}
{\mathop{\rm MSE}\nolimits}  = \sum\limits_{l = 1}^L {\frac{{\left\| {{{{\bf{\hat X}}}_l} - {\bf{X}}} \right\|_F^2}}{{LNR}}},
\end{equation}
where $ {\bf{X}} $ is the true EEG data with \emph{R} channels and each channel has length \emph{N}, $ {{\bf{\hat X}}_l} $ is its estimate in the \emph{l}-th experiment, and \emph{L} is the number of experiments. Both $ {\bf{X}} $ and $ {{\bf{\hat X}}_l} $ are normalized by their Frobenius norms respectively.  When \emph{R} = 1, the matrix \textbf{X} is degenerated into a vector \textbf{x}. In that case, MSE can be used to evaluate single-channel EEG signal reconstruction evaluation. The MSE has variants of other equivalent forms, such as mean L2 error \cite{liu2013multi}, percent of root-mean-square difference (PRD) \cite{abdulghani2012compressive}.
%

Another evaluation function is the mean cross-correlation (MCC). It is equivalent to the Structural SIMilarity index (SSIM), which measures the similarity of two waveforms \cite{wang2009mean} \cite{abdulghani2012compressive} \cite{zhang2013compressed}. It can be formulated as:
\begin{equation}
\label{eq5 MCC}
{\mathop{\rm MCC}\nolimits}  = \sum\limits_{l = 1}^L {\frac{{{\mathop{\rm vec}\nolimits} {{\left( {\bf{X}} \right)}^T}{\mathop{\rm vec}\nolimits} \left( {{{{\bf{\hat X}}}_l}} \right)}}{{L{{\left\| {\bf{X}} \right\|}_F}{{\left\| {{{{\bf{\hat X}}}_l}} \right\|}_F}}}}.
\end{equation}

\subsection{Data material and subjects}
\label{sec5.0}

The used EEG data is the CHB-MIT scalp EEG database which is online available in the \emph{Physiobank} database: http://www.physionet.org/cgi-bin/atm/ATM \cite{shoeb2009application} \cite{goldberger2000physiobank}. Collected at the Children's Hospital Boston, these EEG recordings are from pediatric subjects with intractable seizures. Subjects were monitored without anti-seizure medication in order to characterize their seizures and assess their candidacy for surgical intervention. All the recordings were collected from 22 subjects (5 males, ages 3-22; and 17 females, ages 1.5-19). All used datasets consist of 23-channel EEG recordings which were sampled at 256 samples per second with 16-bit resolution. The international 10-20 system of EEG electrode positions and nomenclature was used for these recordings. More details about the EEG database can be found \cite{shoeb2009application}. In our experiments, the EEG recording $ chb01\_31.edf $ has been selected to demonstrate the recovery algorithms' performance.

In section \ref{sec5.1}, \emph{L} = 500 segments of EEG data are used, i.e. $ {{\mathbf{x}}_l} \in {\mathbb{R}^{N \times 1}},~l = 1,2, \cdots ,L $. They are taken from all the \emph{R}=23 channels sequentially. The length of each segment of the EEG data \textbf{x} is \emph{N} = 256. Each segment of EEG data is normalized by its $ \ell_2 $ norm.

In section \ref{sec5.2}, \emph{L} = 50 segments of 23-channel EEG data are used, i.e. $ {{\mathbf{X}}_l} \in {\mathbb{R}^{N \times R}},~l = 1,2, \cdots ,L $. In each segment of  the EEG data matrix \textbf{X}, the number of sampling points is $ N \times R = 256 \times {\rm{23}}$. Each segment of EEG data is normalized by its Frobenius norm.

\subsection{Single channel EEG signal recovery}
\label{sec5.1}

\begin{figure}[!h]
\centering
\includegraphics[width=0.5\textwidth]{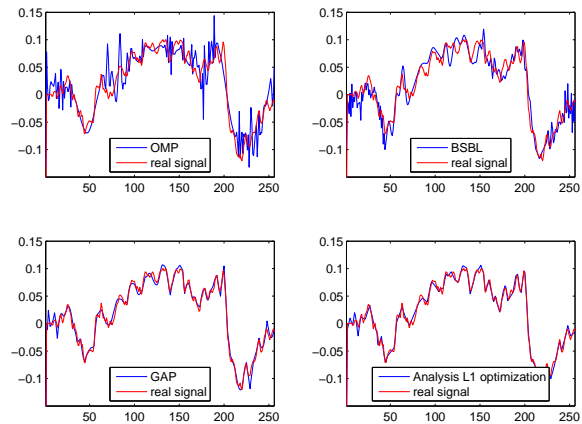}
\caption{EEG signals reconstructed by OMP, BSBL, GAP, and analysis L1 optimization with SSR = 0.35. }
\label{fig101}
\end{figure}

To show how the proposed cosparse signal recovery methods work, we take a segment of single channel EEG signal and reconstruct it from the compressed measurement with SSR = 0.35. The reconstructed and real signals are shown in Fig. \ref{fig101}. We can see that the reconstructed signals from GAP and the analysis L1 optimization methods fit the real signal better than those from the classical OMP and BSBL methods.

\begin{figure}[!h]
        \centering
        ~ 
        \begin{subfigure}[b]{0.5\textwidth}
                \centering
                \includegraphics[width=1\textwidth]{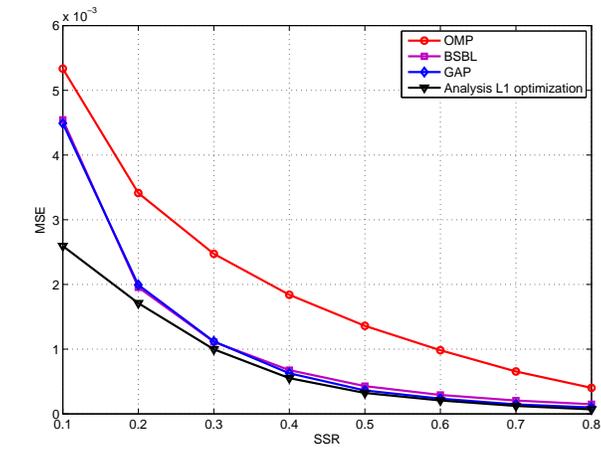}
                \caption{MSE}
                \label{fig6}
        \end{subfigure}
        \qquad
        \centering
        \begin{subfigure}[b]{0.5\textwidth}
                \centering
                \includegraphics[width=1\textwidth]{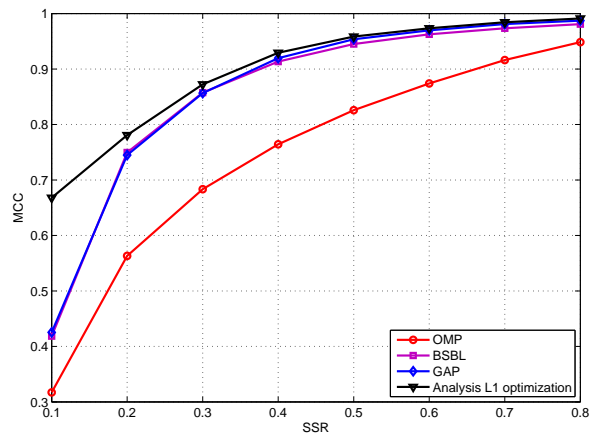}
                \caption{MCC}
                \label{fig7}
        \end{subfigure}%
        \qquad
        ~ 
        \begin{subfigure}[b]{0.5\textwidth}
                \centering
                \includegraphics[width=1\textwidth]{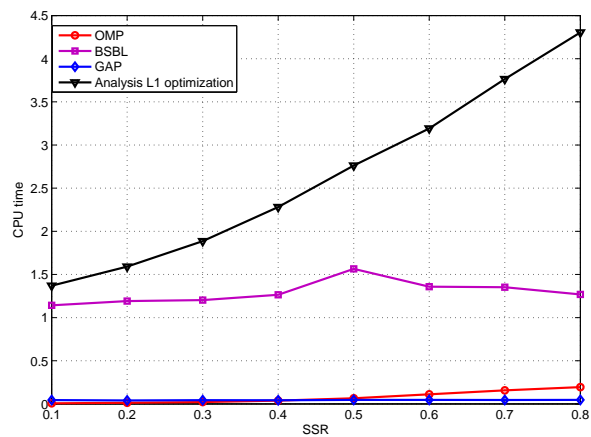}
                \caption{CPU time}
                \label{fig8}
        \end{subfigure}

        \caption{Differences in average performance evaluation of single channel EEG signal recovery from compressed measurements with different SSRs using 500 different single channel EEG segments: (a) MSE vs SSR; (b) MCC vs SSR; (c) CPU Time vs SSR.}\label{figIII}
\end{figure}

Fig. \ref{fig6}, and Fig. \ref{fig7} give the values of MSE and MCC of GAP, OMP, BSBL with different SSRs. We can see that analysis L1 optimization, GAP and BSBL have similar accuracy performance, and they outperform OMP. Analysis L1 optimization is slightly more accurate than GAP, and GAP is slightly more accurate than BSBL. Fig. \ref{fig8} shows that the greedy algorithms GAP and OMP are much faster than BSBL and analysis L1 optimization, and GAP is even slightly faster than OMP. Therefore, if we only care about the accuracy, the analysis L1 optimization is the best choice; and if both accuracy and computational complexity are important, GAP should be a better choice.

\subsection{Multi-channel EEG signal recovery}
\label{sec5.2}

In these experiments, most of the parameters are selected as in section \ref{sec5.1}. Two algorithms for SCLR optimization are used, i.e. interior point method for SCLR optimization and ADMM for SCLR optimization with experienced choices of the parameters $ T_{max} = 5 $, $ \rho = 1 $, $ \eta = 0.05 $. In comparison with the proposed methods, 3 other popular multi-channel sparse / cosparse signal recovery methods are taken too. i.e. BSBL, SOMP and SGAP.

\begin{figure}[!h]
        \centering
        ~ 
        \begin{subfigure}[b]{0.5\textwidth}
                \centering
                \includegraphics[width=1\textwidth]{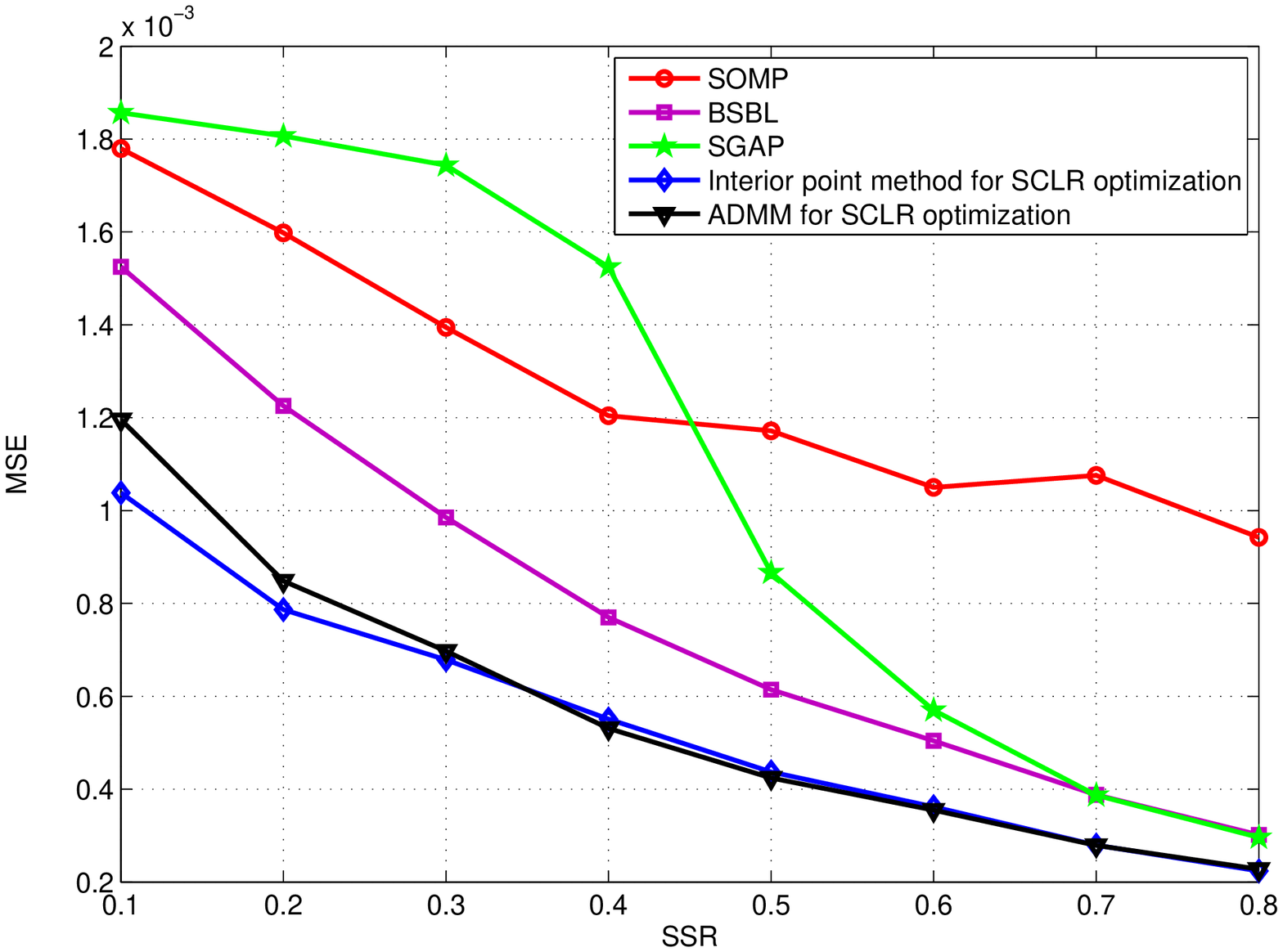}
                \caption{MSE}
                \label{fig10}
        \end{subfigure}
        \qquad
        \centering
        \begin{subfigure}[b]{0.5\textwidth}
                \centering
                \includegraphics[width=1\textwidth]{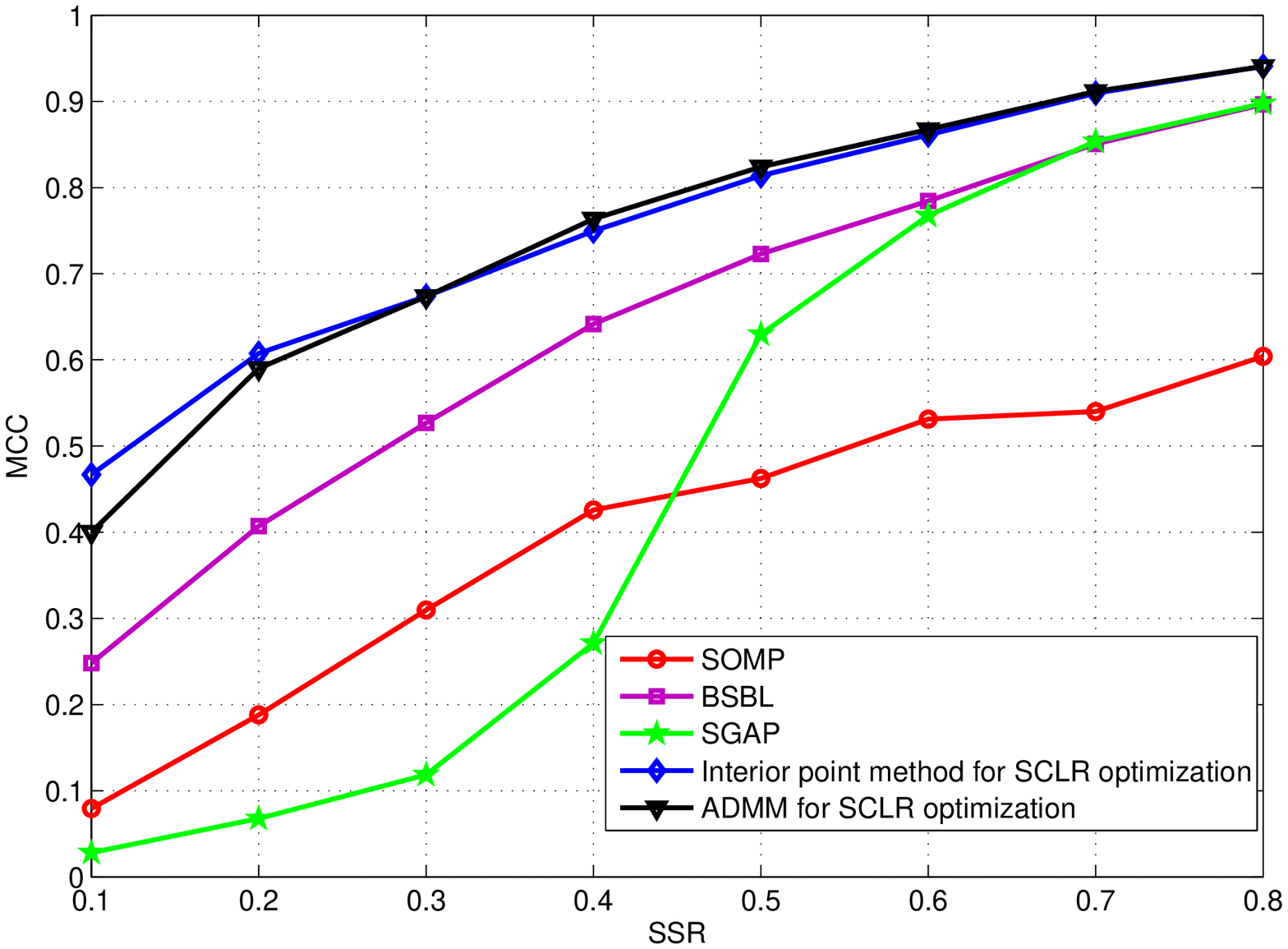}
                \caption{MCC}
                \label{fig11}
        \end{subfigure}%
        \qquad
        ~ 
        \begin{subfigure}[b]{0.5\textwidth}
                \centering
                \includegraphics[width=1\textwidth]{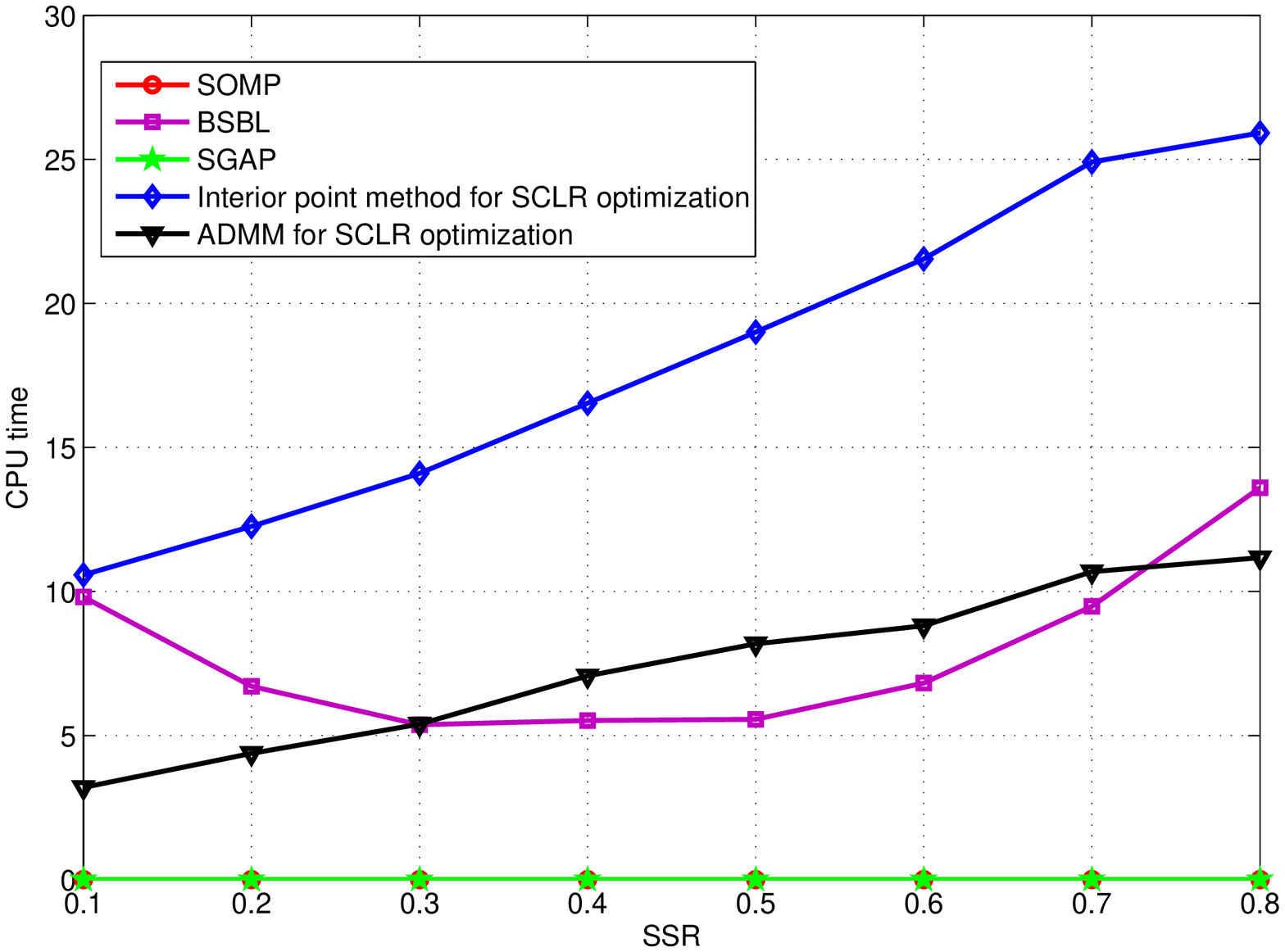}
                \caption{CPU time}
                \label{fig12}
        \end{subfigure}

        \caption{Differences in average performance evaluation of multi-channel EEG signal recovery from compressed measurement with different SSRs using 50 different multi-channel EEG segments: (a) MSE vs SSR; (b) MCC vs SSR; (c) CPU Time vs SSR.}\label{figIV}
\end{figure}

Fig. \ref{fig10}, \ref{fig11} and \ref{fig12} display the values of MSE, MCC and CPU times of the interior point method for the SCLR optimization, ADMM for SCLR optimization, BSBL, SOMP, and SGAP with different values of SSR. We can see that the interior point method for SCLR optimization, ADMM for SCLR optimization have similar accuracy performance, and they outperform the other ones in accuracy. Comparing the speed of these two solutions for SCLR optimization, the ADMM for SCLR optimization is faster. In Fig. \ref{fig12} we can see that the greedy algorithms SOMP and SGAP are much faster than the rest. But their accuracy is much worse and not acceptable. Therefore, we recommend that the ADMM for SCLR optimization should be a better candidate for multi-channel EEG signal recovery than the other methods.

\section{Conclusion}
\label{sec6}

 With the 2nd-order difference matrix as the cosparse analysis dictionary, the EEG signals' cosparsity is exploited for the single-channel EEG signal recovery from compressed measurements. To further enhance the performance, cosparsity and low rank structure are jointly used in the multi-channel EEG signal recovery. In the proposed new optimization model, the $ \ell_0 $ norm constraint is used to encourage cosparsity while Schatten-0 norm constraint is used for low rank structure. To solve the optimization model, two methods are used. One approximates it by relaxing the $ \ell_0 $ and Schatten-0 norms into  $ \ell_1 $ norm and nuclear norm respectively, which leads to a convex optimization. The other way is ADMM which divides the multiple criterion optimization into several connected single criterion optimizations in the form of global consensus optimization. Each single criterion optimization can be solved by a series of existing efficient methods. In numerical experiments, EEG signals' cosparsity for CS is proved by the single-channel EEG data based results; and the multi-channel EEG data results show that the SCLR optimization outperforms all the previous methods.

%
%
%


%
%
%

\begin{thebibliography}{1}





\bibitem{bachmann2012low}
C.~Bachmann et al,
  ``Low-power wireless sensor nodes for ubiquitous long-term biomedical signal
  monitoring,'' \emph{Communications Magazine, IEEE}, vol.~50, no.~1, pp.
  20-27, 2012.

\bibitem{devos2014mobile}
M.~De Vos et al, ``Towards a truly mobile auditory brain¨Ccomputer interface: Exploring the P300 to take away,''
  \emph{ International journal of psychophysiology}, vol.~91, no.~1, pp.46-53, 2014.

\bibitem{debener2012how}
S.~Debener et al, ``How about taking a low-cost, small, and wireless EEG for a walk?,''
  \emph{Psychophysiology}, vol.~49, no.~11, pp.1617-1621, 2012.

\bibitem{abdulghani2012compressive}
A.~M. Abdulghani et al, ``Compressive
  sensing scalp EEG signals: implementations and practical performance,''
  \emph{Medical and biological engineering and computing}, vol.~50, no.~11, pp.
  1137-1145, 2012.

\bibitem{zhang2013compressed}
Z.~Zhang et al, ``Compressed sensing of EEG for
  wireless telemonitoring with low energy consumption and inexpensive
  hardware,'' \emph{Biomedical Engineering, IEEE Transactions on}, vol.~60,
  no.~1, pp. 221-224, 2013.

\bibitem{eldar2012compressed}
Y.~C. Eldar and G.~Kutyniok, \emph{Compressed sensing: theory and
  applications}.\hskip 1em plus 0.5em minus 0.4em\relax Cambridge University
  Press, 2012.

\bibitem{becker2011practical}
S.~R. Becker, ``Practical compressed sensing: modern data acquisition and
  signal processing,'' Ph.D. dissertation, California Institute of Technology, CA,
  2011.

\bibitem{tropp2010computational}
J.~A. Tropp and S.~J. Wright, ``Computational methods for sparse solution of
  linear inverse problems,'' \emph{Proceedings of the IEEE}, vol.~98, no.~6,
  pp. 948-958, 2010.

\bibitem{mamaghanian2011compressed}
H.~Mamaghanian et al, ``Compressed
  sensing for real-time energy-efficient ECG compression on wireless body
  sensor nodes,'' \emph{Biomedical Engineering, IEEE Transactions on}, vol.~58,
  no.~9, pp. 2456-2466, 2011.

\bibitem{nam2013cosparse}
S.~Nam et al, ``The cosparse analysis model
  and algorithms,'' \emph{Applied and Computational Harmonic Analysis},
  vol.~34, no.~1, pp. 30-56, 2013.

\bibitem{candes2011compressed}
E.~J. Candes et al, ``Compressed sensing
  with coherent and redundant dictionaries,'' \emph{Applied and Computational
  Harmonic Analysis}, vol.~31, no.~1, pp. 59-73, 2011.


 \bibitem{peleg2013performance}
T. Peleg and M. Elad,
  ``Performance guarantees of the thresholding algorithm for the cosparse analysis model,'' \emph{Information Theory, IEEE Transactions on}, vol.~59, no.~3, pp. 1832-1845, 2013.


 \bibitem{yan2013an}
C. Yan et al,
  `` An approach of time series piecewise linear representation based on local maximum, minimum and extremum,'' \emph{Journal of Information and Computational Science}, vol.~10, no.~9, pp. 2747-2756, 2013.



\bibitem{durka2005multichannel}
P.~J. Durka et al,
  ``Multichannel matching pursuit and EEG inverse solutions,'' \emph{Journal of
  Neuroscience Methods}, vol. 148, no.~1, pp. 49-59, 2005.

\bibitem{cotter2005sparse}
S.~F. Cotter et al, ``Sparse solutions to
  linear inverse problems with multiple measurement vectors,'' \emph{Signal
  Processing, IEEE Transactions on}, vol.~53, no.~7, pp. 2477-2488, 2005.

\bibitem{zhang2014spatiotemporal}
Z.~Zhang et al, ``Spatiotemporal sparse
  Bayesian learning with applications to compressed sensing of multichannel
  physiological signals,'' \emph{Neural Systems and Rehabilitation Engineering,
  IEEE Transactions on}, vol.~22, no.~6, pp. 1186-1197, 2014.

\bibitem{fauvel2014energy}
S.~Fauvel and R.~K. Ward, ``An energy efficient compressed sensing framework
  for the compression of electroencephalogram signals,'' \emph{Sensors},
  vol.~14, no.~1, pp. 1474-1496, 2014.

\bibitem{srinivasan2013multichannel}
K.~Srinivasan et al, ``Multichannel EEG compression:
  Wavelet-based image and volumetric coding approach,'' \emph{Biomedical and
  Health Informatics, IEEE Journal of}, vol.~17, no.~1, pp. 113-120, 2013.

\bibitem{dauwels2013near}
J.~Dauwels et al, ``Near-lossless
  multichannel EEG compression based on matrix and tensor decompositions,''
  \emph{Biomedical and Health Informatics, IEEE Journal of}, vol.~17, no.~3,
  pp. 708-714, 2013.

\bibitem{candes2013simple}
E.~Cand{\`e}s and B.~Recht, ``Simple bounds for recovering low-complexity
  models,'' \emph{Mathematical Programming}, vol. 141, no. 1-2, pp. 577-589,
  2013.

\bibitem{rohde2011estimation}
A.~Rohde and A.~Tsybakov, ``Estimation of high-dimensional low-rank matrices,''
  \emph{The Annals of Statistics}, vol.~39, no.~2, pp. 887-930, 2011.

\bibitem{recht2010guaranteed}
B.~Recht et al, ``Guaranteed minimum-rank solutions of
  linear matrix equations via nuclear norm minimization,'' \emph{SIAM review},
  vol.~52, no.~3, pp. 471-501, 2010.

\bibitem{boyd2009convex}
S.~Boyd and L.~Vandenberghe, \emph{Convex optimization}.\hskip 1em plus 0.5em
  minus 0.4em\relax Cambridge university press, 2009.



\bibitem{vandenberghe1996semidefinite}
L.~Vandenberghe and S.~Boyd, ``Semidefinite programming,'' \emph{SIAM review},
  vol.~38, no.~1, pp. 49-95, 1996.

\bibitem{grant2012cvx}
M.~Grant et al, ``CVX: Matlab software for disciplined convex
  programming, version 2.0 beta,'' \emph{Recent Advances in Learning and
  Control}, pp. 95-110, 2012.

\bibitem{boyd2011distributed}
S.~Boyd et al, ``Distributed
  optimization and statistical learning via the alternating direction method of
  multipliers,'' \emph{Foundations and Trends{\textregistered} in Machine
  Learning}, vol.~3, no.~1, pp. 1-122, 2011.

\bibitem{eckstein1992douglas}
J.~Eckstein and D.~P. Bertsekas, ``On the Douglas-Rachford splitting method
  and the proximal point algorithm for maximal monotone operators,''
  \emph{Mathematical Programming}, vol.~55, no. 1-3, pp. 293-318, 1992.

\bibitem{tropp2006algorithms}
J.~A. Tropp et al, ``Algorithms for simultaneous
  sparse approximation. part I: Greedy pursuit,'' \emph{Signal Processing},
  vol.~86, no.~3, pp. 572-588, 2006.

\bibitem{avonds2014simultaneous}
Y.~Avonds et al, ``Simultaneous greedy analysis pursuit
  for compressive sensing of multi-channel ecg signals,'' in \emph{Engineering
  in Medicine and Biology Society (EMBC), 2014 Annual International Conference
  of the IEEE}.\hskip 1em plus 0.5em minus 0.4em\relax IEEE, 2014, pp.
  6385-6388.

\bibitem{liu2013multi}
Y.~Liu et al,
  ``Multi-structural signal recovery for biomedical compressive sensing,''
  \emph{Biomedical Engineering, IEEE Transactions on}, vol.~60, no.~10, pp.
  2794-2805, 2013.

\bibitem{wang2009mean}
Z.~Wang and A.~C. Bovik, ``Mean squared error: love it or leave it? a new look
  at signal fidelity measures,'' \emph{Signal Processing Magazine, IEEE},
  vol.~26, no.~1, pp. 98-117, 2009.

\bibitem{shoeb2009application}
A.~H. Shoeb, ``Application of machine learning to epileptic seizure onset
  detection and treatment,'' Ph.D. dissertation, Massachusetts Institute of
  Technology, 2009.

\bibitem{goldberger2000physiobank}
A.~L. Goldberger et al, ``Physiobank,
  physiotoolkit, and physionet components of a new research resource for
  complex physiologic signals,'' \emph{Circulation}, vol. 101, no.~23, pp.
  e215-e220, 2000.




\end{thebibliography}
%

\end{document}